# Investigation of the Influence of a field-free electrostatic Potential on the Electron Mass with Barkhausen-Kurz Oscillation

M. Weikert and M. Tajmar[1]

Institute of Aerospace Engineering, Technische Universität Dresden, 01062 Dresden, Germany

**Abstract**

According to Weber's electrodynamics, Assis showed analytically, that a field-free electrostatic potential delivered by a spherical shell causes a force upon a moving electrical charge in the center of that shell. This force can be interpreted as a result of the change in inertial mass of the charge. In order to prove this theory, Mikhailov published two type of experimental setups: One using vacuum cathode tube and another using glow-discharge-lamps to generate oscillating and accelerating electrons. Whereas the glow-discharge experiment was already evaluated by several groups, here we are focusing on replicating the vacuum tube configuration. Under right circumstances, electrons inside a vacuum tube start to oscillate around a grid electrode, which is called Barkhausen-Kurz oscillations. However, we found that Mikhailov's setup does not produce these kind of oscillations and therefore the theory that he applied in the interpretation of his measurements is not correct. We succeeded in generating Barkhausen-Kurz oscillations with a different vacuum tube and found no frequency shifts below an order of magnitude of Assis's prediction by operating the tube inside a charged spherical shell that would indicate a change in the electron's mass. However, since both the mass as well as the geometry factor of the electron cloud contribute to the oscillator frequency, we believe that this setup is not suitable to investigate Weber-type electrodynamic effects.

---

[1] Professor, Email: martin.tajmar@tu-dresden.de

## 1. Introduction

In the 19$^{th}$ century, Wilhelm Eduard Weber proposed a mathematical formulation of the electrodynamic effects he observed by performing multiple precise measurements. As a result, he presented a formulation extending Coulomb's law by including higher derivatives of the distance between its two charges $r$, i.e. the speed and acceleration between these charges,

$$F = \frac{q_1 q_2}{4\pi\varepsilon_0} \frac{\hat{r}}{r^2} \cdot \left(1 - \frac{\dot{r}^2}{2c^2} + \frac{r\ddot{r}}{c^2}\right) \tag{1}$$

It can be shown, that this formulation is able to represent all effects described by Maxwell equations, the Lorentz force law as well as other effects such as an apparent change of the inertial mass of a charged moving inside a charged sphere [1]–[3]. According to Assis [4], the inertial mass of an accelerated charge $q$ inside a charged spherical shell (with a total charge $Q$ for the shell) with a radius $R$, permittivity $\varepsilon_0$, and a field-free potential $\Phi$ can be expressed as

$$\Delta m = \frac{qQ}{12\pi\varepsilon_0 c^2 R} = \frac{q\Phi}{3c^2} \tag{2}$$

It is important to note that this accelerated charge must be free of influence of the atomic grid of a conductor to be valid.

Towards the end of the 19$^{th}$ century, both Maxwell's and Weber's theory were both well known. However, the consequence of a variable mass was then thought to be unphysical and Weber's theory was dismissed without testing if such a consequence is real. However, the concept of a variable charged mass inside an electric potential is not too surprising as this is simply a consequence of Einstein's E=mc² and the electrostatic potential energy [5], [6]. However, there has been a strong debate if electrostatic potential energy indeed contributed to the inertial mass of the charge itself or to the overall system [7].

It took some 100 years after Weber's theory was forgotten until Mikhailov published an experiment where he claimed to have observed just this effect [8], [9]. He used a Neon-discharge lamp oscillator inside a charged Faraday-cage and observed a frequency shift proportional to the applied potential on the cage. His measurements compared well to Weber's and Assis's prediction. Later, he published another experiment using Barkhausen-Kurz oscillations inside a vacuum tube [10] where he claimed again to have seen frequency shifts according Weber's theory. Shortly afterwards, Junginger and Popovic [11] repeated Mikhailov's first Neon-lamp discharge experiment with a refined experimental setup and found no effect which again closed the book on Weber's theory.

However, there is a wealth of recent analysis and experiments that challenge this quick dismissal because the implications would be indeed far-reaching. Lörincz and Tajmar have recently shown that the overall concept of Mikhailov's experiment was flawed [12]. A Neon-discharge lamp is not at all representative to an isolated charged inside a charged Faraday cage that Assis used for his analysis due to the contributions from the surrounding ions in the discharge. Therefore, Junginger and Popovic's results does not at all invalidate Weber's theory. Moreover, a number of recent experiments with electromagnetic induction and electron beam deflections in magnetic fields also seem to favor Weber's over Maxwell's predictions [13]–[15]. Also Assis has shown that self-induction in a circuit can also be successfully modelled using Weber's electrodynamics and the variable mass term [16].

In this paper, we will evaluate Mikhailov's second experiment using Barkhausen-Kurz oscillations, which is a much cleaner approach than the Neon-lamp discharge experiment due to the lack of surrounding ions. We also used a much better experimental setup and measurement equipment, which allowed us to carefully assess if this experiment is indeed suitable to see a Weber like mass variation.

## 1.1 Oscillating electrons

One way of measuring the mass of an electron is to look at its acceleration behavior. While time-of-flight-measurements are technically challenging for a small number of charges, the measurement of the signal of an electromagnetic oscillation is quite easy. So, the linear movement of an electron accelerated between electrodes can be transformed in an oscillation between, respectively around, electrodes. A known technical solution is a so-called Barkhausen-Kurz oscillation [17]. Here, electrons are emitted by a heated cathode and accelerated towards a central grid-anode. Most electrons pass the grid in direction of another cathode. There they are decelerated and accelerated back to the central grid-anode (Fig. 1). Using the right accelerating voltage, an oscillation forms without an external resonator. The frequency of this oscillation is adjustable within a specific voltage range. This effect can be easily produced inside a vacuum cathode triode tube. Depending on the accelerating voltage $U$, the geometric parameter $d$ (distance from cathode to grid-anode) of the tube, the electrical charge of the electron $e$ and electron mass $m$, frequencies from hundreds of megahertz up to a few gigahertz are achievable as shown by

$$f \cong \sqrt{\frac{eU}{2m}} \cdot \frac{1}{d} \qquad (3)$$

The frequency of the Barkhausen-Kurz oscillation can be directly measured by using Leech lines or from the analysis of the electro-magnetic signal emitted by a feed-line connected to the grid-anode. According to Equs. (2) and (3), the emitted frequency under the influence of a high-potential spherical shell should lead to a new frequency Weber-frequency $f_W$

$$f_W = \frac{f_0}{\sqrt{1 - \frac{m_W}{m_0}}} = \frac{f_0}{\sqrt{1 - \frac{q \cdot \Phi}{3c^2 m_0}}} \qquad (4)$$

Therefore, a change of the emitted frequency under the influence of a field-free potential should occur as shown in Table 1. It is important to note that Equ. (3) is only an approximate relationship which assumes a fixed geometrical parameter $d$. However, this geometric parameter describes the length scale of the electron cloud, which oscillates around the grid. In case the inertial mass of the electron becomes smaller, the electron cloud radius may become larger (due to the reduced inertia) which could counterbalance the effect we are looking for.

## 2. Barkhausen Experiment: Mikhailov's Setup

To provide free accelerated electrons without disturbance of an electrical conductor, Mikhailov used a Russian-made tetrode vacuum tube (6Э5П) in a circuit with two capacitors (24pF) and two resistors (12kΩ and 1.3MΩ) as shown in Fig. 2. The λ/4-transmission-line antenna was connected to 108V from a DC-power-supply and emitted a 256MHz sinusoidal signal. This signal-generator was placed inside an Indium-Gallium-coated spherical shell, which was charged with up to 1.25kV. Mikhailov detected the emitted signal with a receiver and a microampere meter. He assumed that the current is directly proportional to the emitted frequency without measuring it directly (probably due to lack of available equipment). He observed a linear relationship between the potential of the spherical shell and the current induced into the receiver. He then concluded that a change in the emitted frequency resulted from a change in the electron mass.

However, his experimental setup already raises concerns about the actual nature of the oscillations that he has claimed to have observed. It is likely, that Mikhailov's oscillation is a result of the continuously changing polarity of the anode and one of the grids of his tube, caused by loading and unloading of the used capacitors. Additionally, the use of capacitors and resistors, which may constitute an external resonating circuit, contradict Barkhausen's theory.

Mikhailov used a receiver, which was adjusted to one specific frequency and a microampere meter showing the induced current. From a changing induced current, he inferred a change in the emitted frequency caused by a change in the mass of the oscillating electrons. Let us now consider the emitted signal as a power curve (Fig. 3). Mikhailovs receiver measured the power of a specific frequency his receiver was adjusted to. If the true emitted signal (orange curve) was of a lower frequency than Mikahilov's expected signal (red line) and the generated frequency increased, he would have seen an increase in power and current with the amperemeter. If Mikahilov's expected frequency (green line) was too low, he would have seen a decrease in current with increasing emitted frequency. In this way, it can be shown, that the experimental setting of Mikhailov was not suitable for proofing Assis theory on changing electron mass according to Webers electrodynamics.

## 3. Replication of Mikhailov's Set-up

To verify Mikhailov's experimental results, his setup was reproduced and measured using a spectrum analyzer. This allows the direct measurement of the electrical power for a wide range of emitted frequencies in real time. Thus, changes in frequency can be observed immediately.

Instead of Mikhailov's 24 pF, a slightly different 22pF capacitor had to be used as this was available for the expected voltages. The vacuum tube 6Э5П was placed inside a 3D-printed multi-layer shell (Fig. 4), which was lined up with Aluminum foil. Dimensions were chosen to ensure a secure use with a potential of up to 15kV. The multi-layer-design allowed to test the potential with and without the influence of a ground and without the risk of flashovers in electronics, low-voltage-supply or to the surroundings. The used electric components (capacitors & resistors on a circuit board and emitter-antenna) were placed outside the shell (Fig. 5). Thus, it was assumed that the field-free potential would only interact with the accelerating electrons in the vacuum tube.

First, the working range of the modified design had to be evaluated using a spectrum analyzer. At Mikhailov's working voltage of 108V, no signal was picked up by our instruments. We had to lower the voltage to 36V where we could then see a signal at 114MHz. In contradiction to a Barkhausen-Kurz oscillation, this frequency was not dependent on the used voltage. A change in voltage did not lead to a change in frequency, but the signal collapsed outside a certain voltage range. The strong discrepancy between the emitted frequency and the expected frequency is supposedly caused by the different total capacity of the experimental setting. The age and condition of the vacuum tube might be another reason for the described discrepancies.

For experiments with the high-potential spherical shell, four 9V-batteries were used as power supply for the tube. Cathode heating was powered by a plug-AC/DC-transformer.

The spherical shell was designed and printed as 3 pairs of half-shells (Fig. 4) enabling the use of two separate aluminum-foil layers. Dimensions and material thickness were chosen to ensure electrical isolation and avoid flash-overs in electronics, power supplies and the surroundings. The vacuum tube was placed near the shell center and fed through a narrow channel, which means that the spherical shell was not closed all around.

First measurements with the inner shell set to high potential of several kV and a grounded outer shell led to an unstable emitted frequency. Therefore, the outer shell and outer aluminum-layer were abandoned.

Our measurements used an increasing potential from 0kV to +10kV and from 0kV to -10kV in 1kV steps as shown in Fig. 6. The first run from 0kV to +10kV showed a decrease of the emitted frequency from 108.46MHz to 108.37Mhz. The second run from 0kV to -10kV showed a decrease from 108.34MHz to 108.3MHz. This effect was identified as a thermal drift of the vacuum tube, which was completed after 30 minutes of heating. Therefore, subsequent measurements were taken after a heating time of 40 minutes.

Considering the heating time, measurements were made resetting the potential after each 1kV step back to 0kV. In this way, at least three runs in the potential range from 0kV to +10kV and 0kV to -10kV were made. No significant change in emitted frequency was observed in the expected magnitude as shown in Fig. 7. Due to the spectrum analyzer and the much higher voltage used as compared to Mikhailov, our measurement resolution was at least one order of magnitude better (factor 50) than required to see the effect according to Weber's theory.

## 4. Barkhausen Experiment: TU Dresden Setup

It has been shown that Mikhailov's results were not reproducible. Nevertheless, considering the used circuit design, the emitted frequency and the measurement method, it is not possible to invalidate Assis' theory because Mikhailov did not fulfill basic requirements. Therefore, it was necessary to ensure the generation of a Barkhausen-Kurz-oscillation, to measure the generated frequency as accurately as possible and eliminate or reduce the impacts of side effects (e.g. tube heating).

While properly connected, a Barkhausen-Kurz-oscillation was not successfully generated with the 6Э5П vacuum tetrode. Therefore, a PC86 vacuum triode was used. Here, an acceleration voltage of 3.8V and a current of 25mA (although the operating limit of the tube was 20mA) produced an electromagnetic signal of 2.15GHz with a power of -72dbm. The emitted frequency was clearly dependent on the square root of the supply voltage within our operating range. This is a convincing evidence, that a Barkhausen-Kurz-oscillation was generated. The tube was also placed in the center of the 3D-printed multi-layer shell. The feedline was directly connected to the grid anode and was used as signal emitter (Fig. 8). Measuring of the emitted frequency was also done by a spectrum analyzer.

Using an adjustable DC-power-supply for operating voltage and a plug-AC/DC-transformer for heating, measurements were made with an operating voltage of 3.8V, 4.0V, and 4.2V. As before, a potential of 0kV up to +10kV and 0kV up to -10kV was used increasing in 1kV-steps and returning to 0kV after each measurement. The potential was connected to a single aluminum layer that acted as the charged Faraday cage around our vacuum tube.

Respecting the heating time, three runs of measurement were made for three operating voltages and therefore three basic frequencies with positive and negative potential. However, no change in emitted frequency was observed within our measurement resolution as shown in Fig. 9. Again the noise of our frequency shift was below the expected value at 10 kV by a factor of 50.

Using three different power-supplies for operating the tube, heating the cathode and powering the potential shell, a possible effect due to three different ground-potentials was considered. This was eliminated by using batteries for tube-operating and heating, and coupling them to a common potential with the potential shell. There was no measurable effect for a potential up to 12kV.

A remarkable effect occurred when separating the heating battery. While increasing the potential, an increase of the emitted frequency was observable. Reaching a stationary potential, the frequency got back to the original value. The effect was stronger with an increase in the positive potential range. The increase of negative potential (from -12kV to 0kV) produced a weaker increase in frequency. We interpret this effect due to the change in frequency because of additional electron acceleration due to the potential difference between the heating wire and the cathode.

## 5. Conclusions and Discussion

None of the experimental setups could reproduce the results claimed in Mikhailov's paper [10]. We saw that Mikhailov did not even generate Barkhausen-Kurz oscillations, which we were able to get using a different vacuum tube. However, our tests show that no frequency change effect was seen within our measurement resolution by a factor of 50. Nevertheless, this result is not sufficient to invalidate Assis's prediction based on Weber's theory. As outlined in section 1.1, the frequency is depending on both the electron's mass as well as the size of the electron cloud, which is also influences by the electron mass. A smaller electron mass should increase the frequency and make the electron cloud larger. According to Equ. (3), both effects would influence the frequency shift in two opposite directions and therefore the effect we were looking far may well have been hidden by not knowing exactly the size of the actual electron cloud.

It seems that also the second approach (first Neon-lamp discharge, then Barkhausen-Kurz oscillations) from Mikhailov is not suitable to look for new effects from Weber-type electrodynamics. New experimental approaches such as electron beam deflections in magnetic fields [15] or time-of-flight measurements seem much more promising.

## Acknowledgement

We would like to thank A.K.T. Assis for many valuable discussions.

| Potential | Frequency Change |
|---|---|
| 1000kV | -22.2% |
| 100kV | -3.11% |
| 10kV | -0.33% |
| 1kV | -0.03% |
| 0kV | 0% |
| -1kV | +0.03% |
| -10kV | +0.33% |
| -100kV | +3.43% |
| -1000kV | +69.59% |

*Table 1: Calculated Frequency Change under Influence of a Field-Free Potential according to Assis* [4]

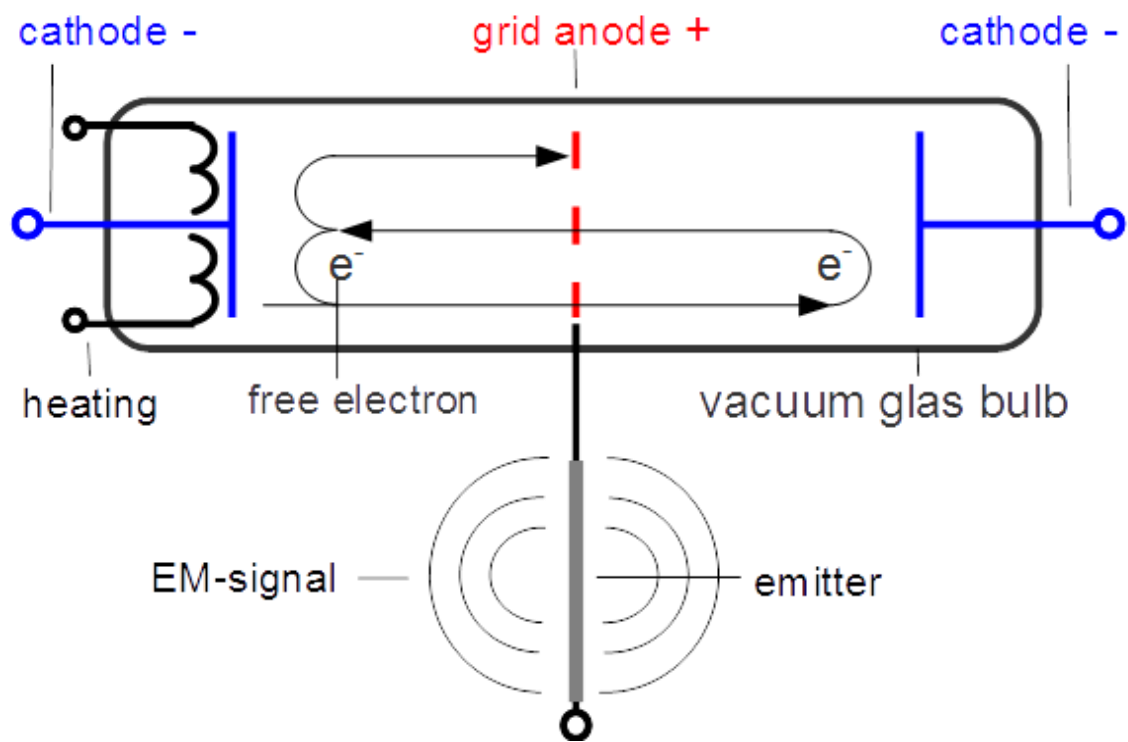

*Figure 1: Functional Principle of a Barkhausen-Kurz-Oscillator*

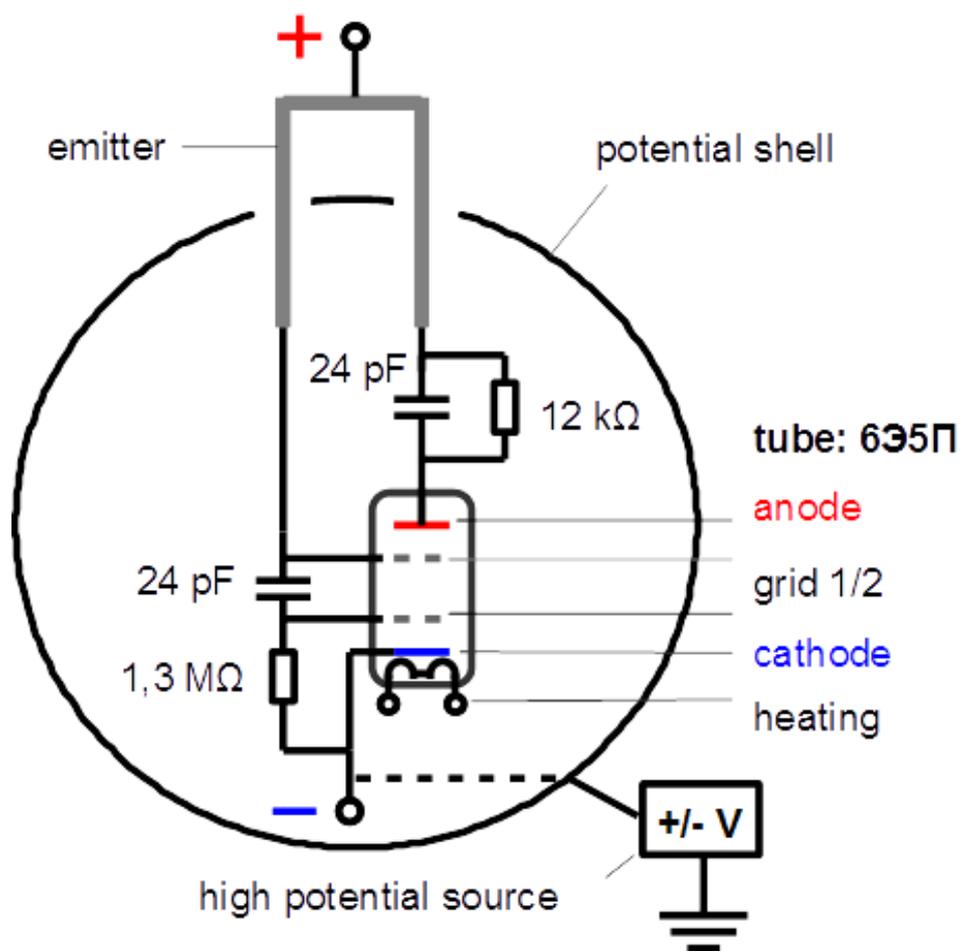

*Figure 2: Layout of Mikhailov's Experimental Setup* [10]

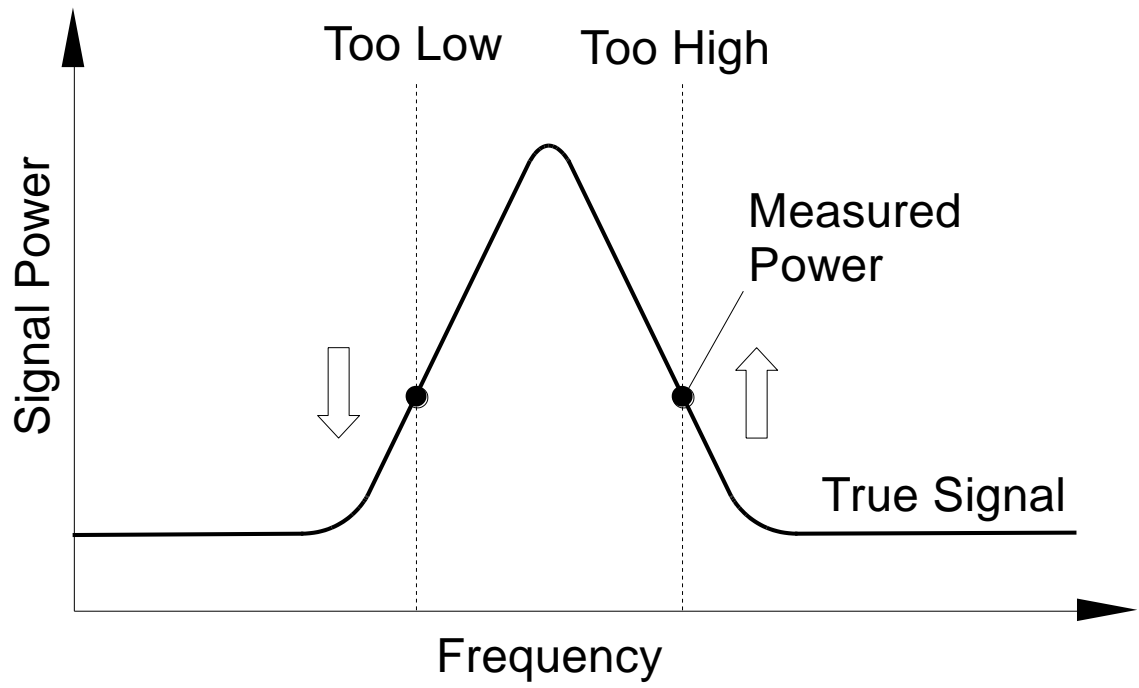

*Figure 3: Power Representation of a Measured Signal relating to the True Signal*

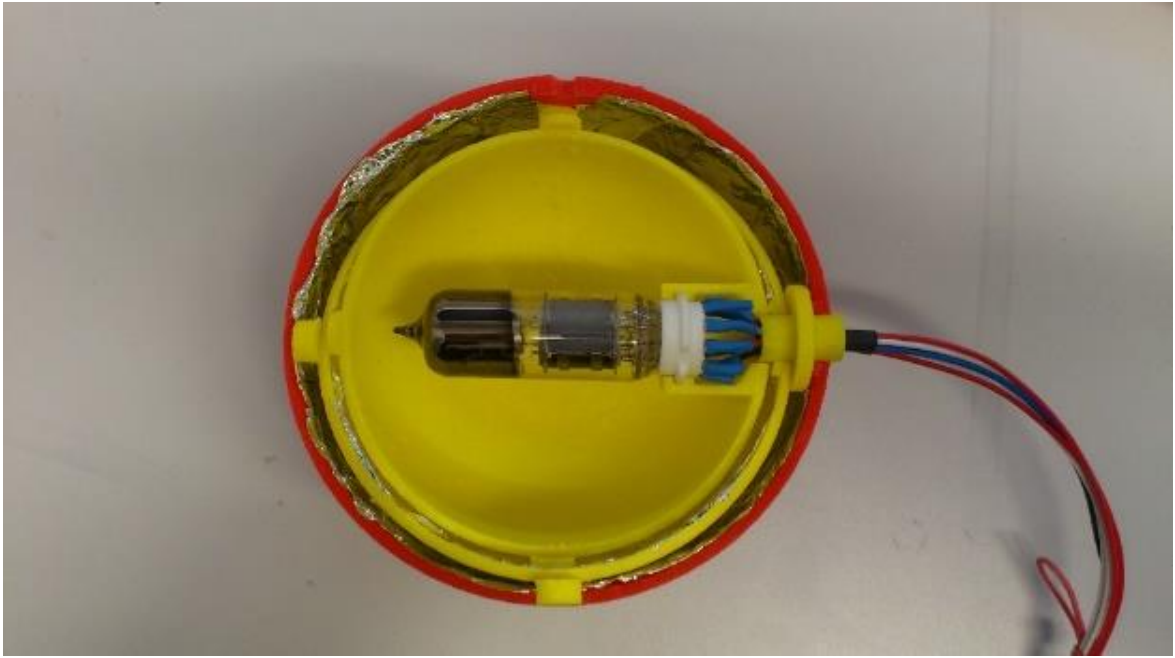

*Figure 4: Vacuum Tube 6Э5П in 3D-printed, Aluminum lined Multi-layer Spherical Shell* [18]

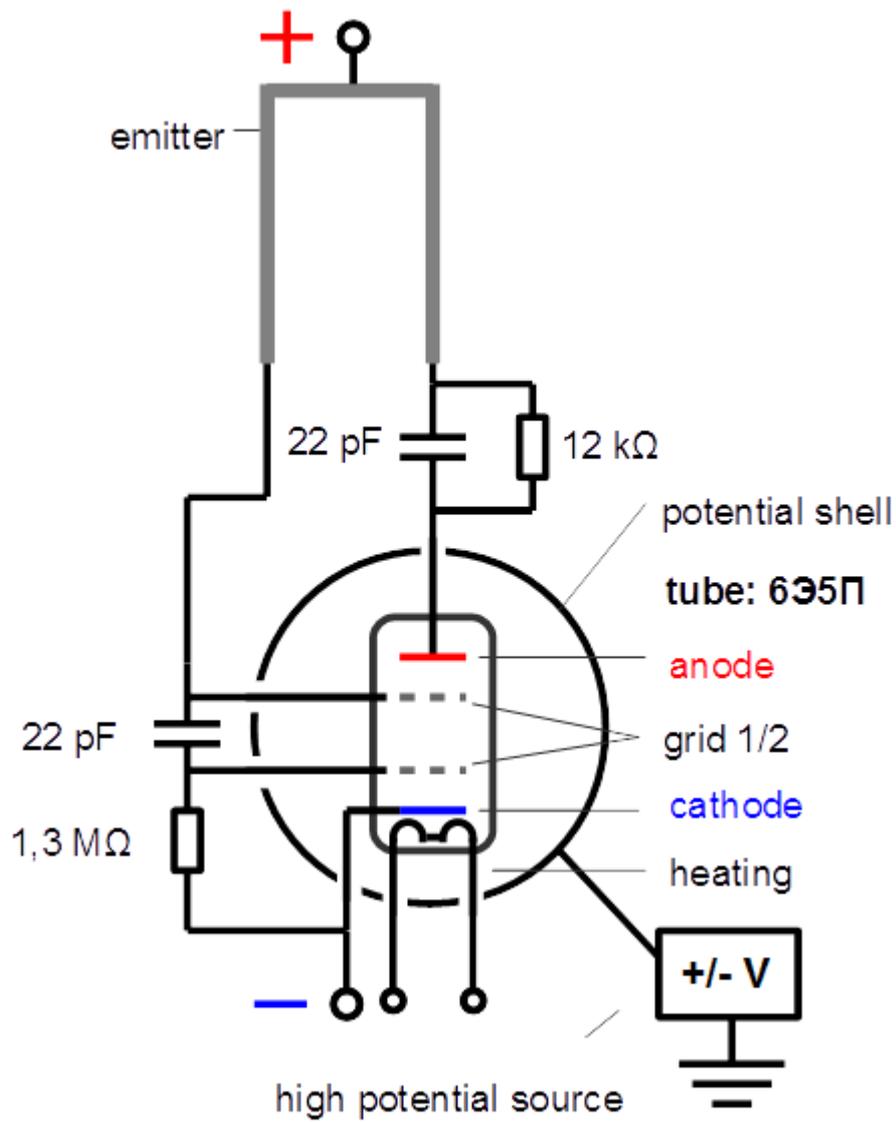

*Figure 5: Layout of the Reproduction of Mikhailov's setup*

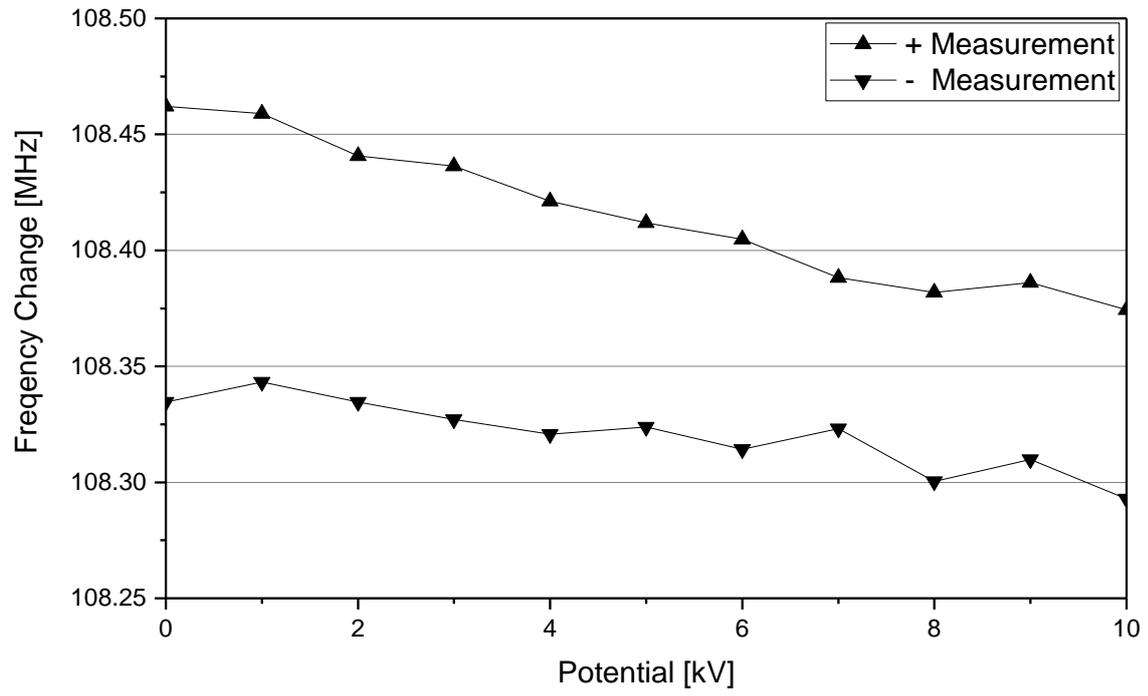

*Figure 6: Seemingly Change of Frequency while Changing Potential in Mikhailov's Setup*

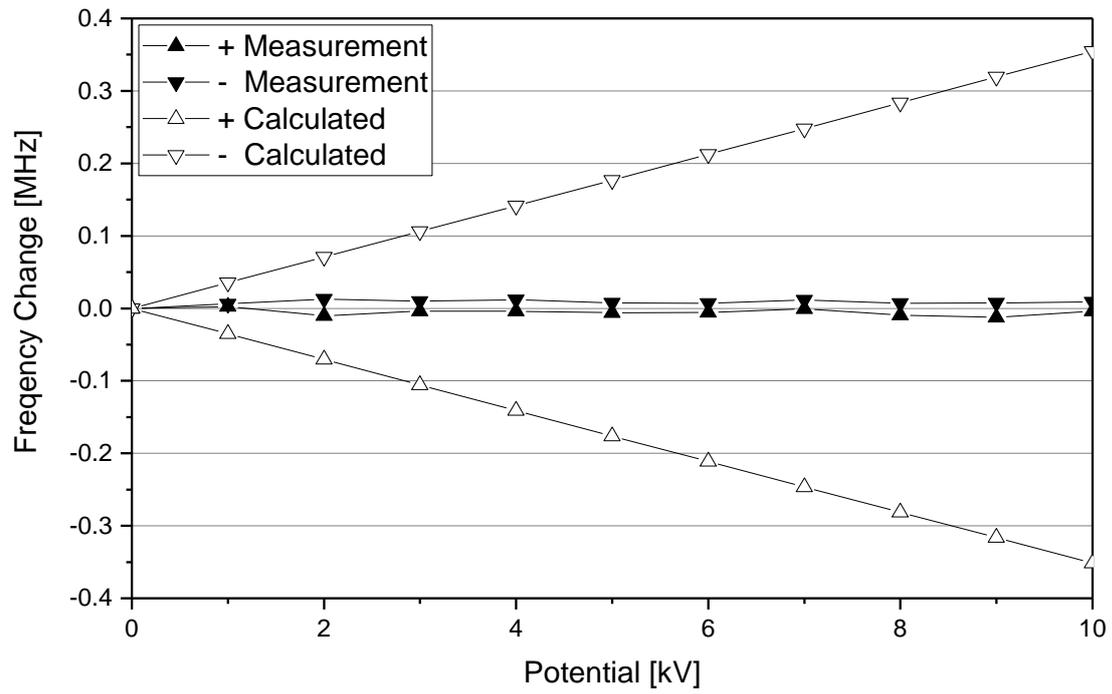

*Figure 7: Comparison of Calculated and Measured Change of Frequency under Influence of a Field-Free Electrostatic Potential in Mikhailov's Setup*

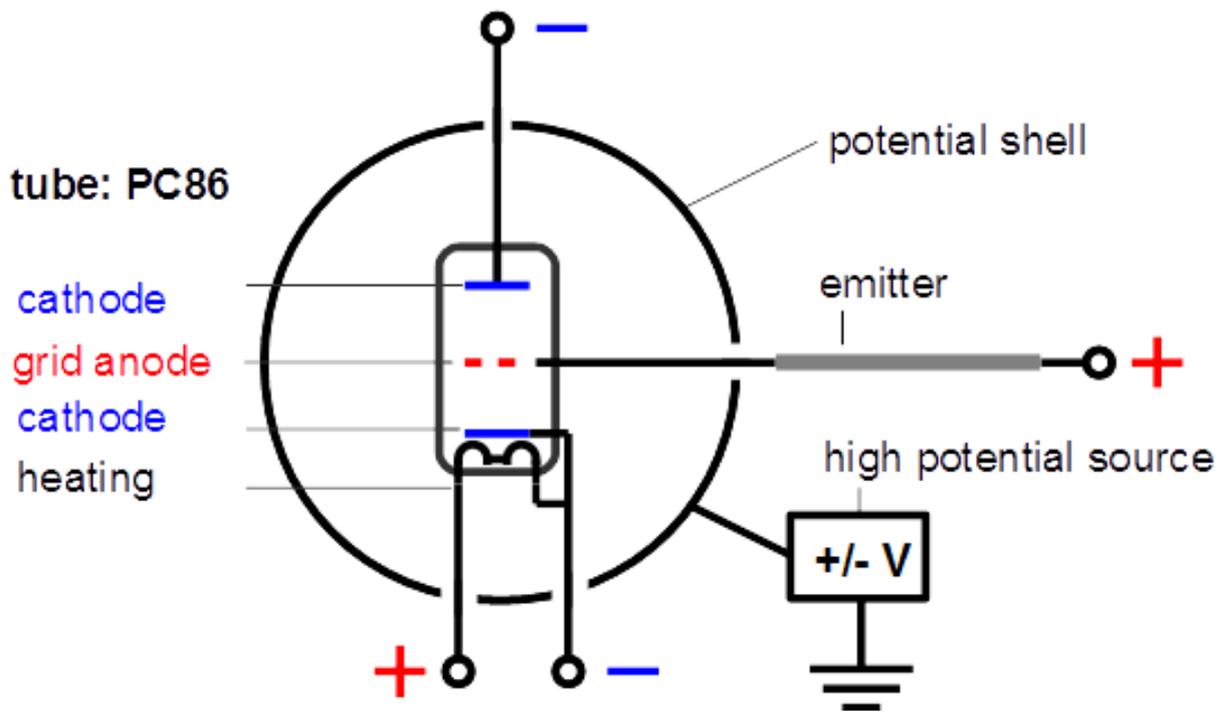

Figure 8: Layout of the Setup using a Barkhausen-Kurz-Oscillation

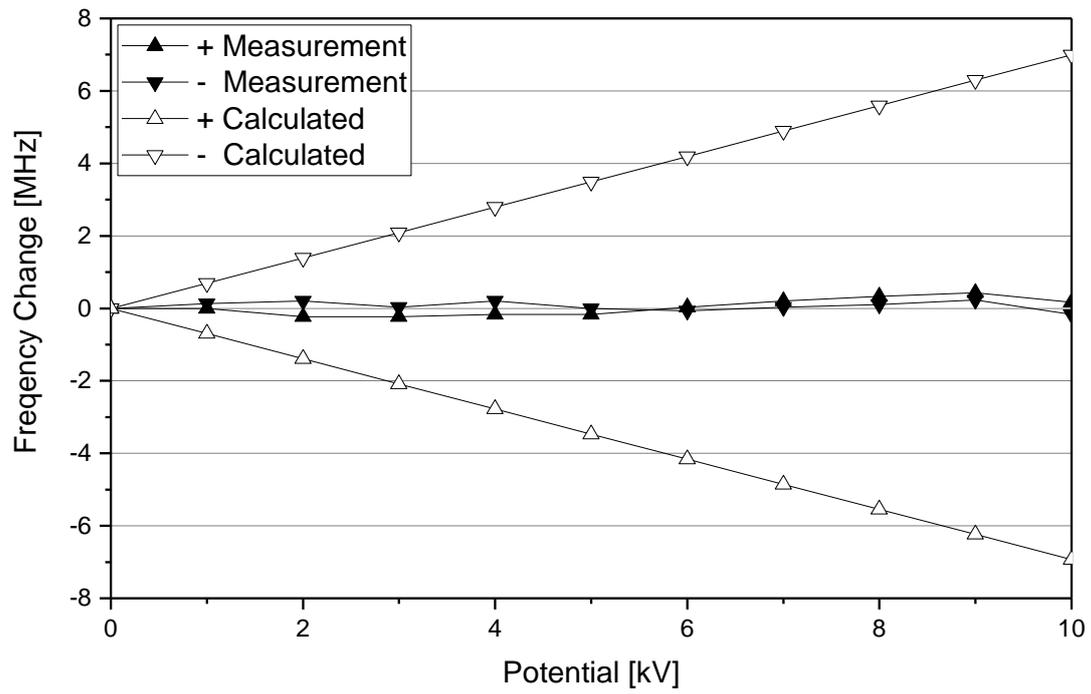

*Figure 9: Comparison of Calculated and Measured Change of Frequency under Influence of a Field-Free Electrostatic Potential in TUD Setup using a PC86 Vacuum Triode*